\documentclass[prl,superscriptaddress,twocolumn]{revtex4-1}
\usepackage[latin9]{inputenc}
\usepackage{amsmath}
\usepackage{amssymb}
\usepackage{amsthm}
\usepackage{graphicx}
\usepackage{color}
\usepackage[margin=1in]{geometry}

\def\identity{\leavevmode\hbox{\small1\kern-3.8pt\normalsize1}}

\newcommand{\ket}[1]{\left | #1 \right\rangle}
\newcommand{\bra}[1]{\left \langle #1 \right |}

\renewcommand{\epsilon}{\varepsilon}

\newcommand{\comm}[1]{\left[#1\right]}
\newcommand{\abs}[1]{\left|#1\right|}
\renewcommand{\O}[1]{O\left(#1\right)}
\usepackage{color}

\usepackage[ruled,linesnumbered]{algorithm2e}
\SetAlgorithmName{Protocol}{protocol}{List of Protocols}
\usepackage{ragged2e}
\usepackage{setspace}

\renewcommand{\section}[1]{\emph{#1}-- }
\newcommand\jus[1]{\medskip\par\noindent\textbf{}\justifying} 
\begin{document}

\title{Blind quantum computation using the central spin Hamiltonian}
\author{Minh~Cong~Tran}
\affiliation{Joint Center for Quantum Information and Computer Science, NIST/University of Maryland, College Park, Maryland 20742, USA}
\affiliation{Joint Quantum Institute, NIST/University of Maryland, College Park, Maryland 20742, USA}

\author{Jacob~M.~Taylor}
\affiliation{Joint Center for Quantum Information and Computer Science, NIST/University of Maryland, College Park, Maryland 20742, USA}
\affiliation{Joint Quantum Institute, NIST/University of Maryland, College Park, Maryland 20742, USA}
\affiliation{Research Center for Advanced Science and Technology,
	University of Tokyo, Meguro-ku, Tokyo 153-8904, Japan}

\begin{abstract}
Blindness is a desirable feature in delegated computation. In the classical setting, blind computations protect the data or even the program run by a server. 
In the quantum regime, blind computing may also enable testing computational or other quantum properties of the server system.
Here we propose a scheme for universal blind quantum computation using a quantum simulator capable of emulating Heisenberg-like Hamiltonians.
Our scheme is inspired by the central spin Hamiltonian in which a single spin controls dynamics of a number of bath spins.
We show how, by manipulating this spin, a client that only accesses the central spin can effectively perform blind computation on the bath spins.
Remarkably, two-way quantum communication mediated by the central spin is sufficient to ensure security in the scheme.
Finally, we provide explicit examples of how our universal blind quantum computation enables verification of the power of the server from classical to stabilizer to full BQP computation.
\end{abstract}

\maketitle


Tremendous progress in the implementation of quantum simulators~\cite{feynman82,houck12,ma14,omalley16,hensgens17,loredo16,monroe17,lukin17} has led us to an enviable scenario in which predicting the general dynamics of a quantum simulator exceeds available classical computation power. 
How then are we to measure the performance of such devices?
In benchmarking, one technique is so-called homomorphic encryption in which the desired computation is hidden from, e.g., the server upon which it is implemented and then decrypted post facto. 
In the quantum domain, such blind computing has been suggested using the circuit model and the measurement-based approaches~\cite{Childs05,Broadbent09}.
These protocols rely upon a high bandwidth quantum communication channel between the server and the client. 
Further developments along these lines have improved security, blindness, and provided a connection to quantum interactive proofs~\cite{Dunjko12,Morimae12,Dunjko14,Morimae15,Aharonov10,Aharonov17,Fitzsimons17,Gheorghiu17}.		

Inspired by the central spin Hamiltonian~\cite{Gaudin76,Taylor03,Yuzbashyan05}, here we build from the ground up a blind computation scheme using a quantum simulator. While we require two-way quantum communication, we show that passing a single qubit back and forth suffices to ensure security. 
This approach should be accessible not only in natural central spin implementation, such as quantum dots~\cite{Merkulov2002,Khaetskii2002,deSousa2003-dots,Schliemann03,Erlingsson04,Chekhovich2013}, NV centers~\cite{Wrachtrup06,Childress06,Balasubramanian09,Shin13,Wang13,Hall14} and NMR molecules~\cite{Pastawski95,Cory98,Laflamme2002}, but also in simulators that can implement Heisenberg-like interactions such as ion traps~\cite{Porras04,Lanyon11,Arrazola16,Bermudez17} and circuit QED systems~\cite{Cho08,houck12,Heras14,Salathe15,Lamata17}.

The paper is structured as follows. 
We first review the central spin Hamiltonian where a central spin is coupled to a number of bath spins. 
We show how different states of the central spin effectively leads to different dynamics of the bath spins. 
With this observation, we present a delegated simulation scheme in which the dynamics of the bath spins are controlled by a single central spin communicated back and forth between the client and the server.
However, the simulation is blind only if the server is trusted not to measure the central spin.
Such a vulnerability is later removed in an improved scheme with ``honeypots" added to detect measurement attempts during the computation. 
Finally, we show that our blind simulation scheme is capable of simulating a universal quantum gate set, and therefore allows the client to perform universal, blind computation on the server.


\section{Central spin model}
We consider a system in which the central spin Hamiltonian is either natural (e.g. in quantum dots, NV centers)  or can be simulated (e.g. using ion traps and circuit QED).
Such a Hamiltonian describes interaction between a spin-$\frac12$ central spin $\vec S_0$ with $n$ spin-$\frac{1}{2}$ bath spins $\vec S_j$ for $j=1,\dots,n$,
\begin{align}
	H_c = \sum_{j=1}^n \gamma_j \vec S_0\cdot \vec S_j - h_0 S_0^z - \sum_{j=1}^n h_j S_j^z, \label{EQ_HC}
\end{align} 
where $\vec S_j=\left(S_j^x,S_j^y,S_j^z\right)$ is the spin operator vector of the $j$th spin and $\gamma_j$ denotes the interaction strength between $\vec S_0$ and $\vec S_j$.
The last two terms are the result of the interaction between the spins and an external magnetic field. 
Without loss of generality, we assume the field to be along the $z$ axis. 
For our blind computation protocol, we also assume that $h_0$ and $h_j,\gamma_j$ for $j=1,\dots,n$ are tunable parameters of the system.

In the following discussion, we further assume that the magnetic field on the central spin is much larger than the interaction between the spins, i.e. $h_0\gg n\gamma = \eta$ with $\gamma = \frac{1}{n}\sum_j\abs{\gamma_j}$ being the average interaction strength. 
Without loss of generality and for simplicity, we further set $h_0 = 1$ in our calculation. 
In this $\eta\ll 1$ limit, the Hilbert space is well separated into two subspaces, each corresponds to an eigenstate of $S_0^z$ of the central spin. 
Although there is no interaction term between the bath spins in Eq.~\eqref{EQ_HC}, they can still interact with each other via an interaction mediated by the central spin~\cite{YaoLS06,LiuYS07}. 

Using the Schrieffer-Wolff approximation~\cite{Bravyi11}, we find the Hamiltonians describing such effective interactions among the bath spins in the two subspaces (Supplemental Material). 
For example, in the subspace that corresponds to the central spin being in $\ket{0}$, i.e. the ``up" state, the effective Hamiltonian is
\begin{align}
	H_{\uparrow} = &-\frac{1}{2} \sum_{j<k}\gamma_j\gamma_k\left(S_j^xS_k^x+S_j^yS_k^y\right)\nonumber\\
	&-\sum_{j=1}^n\left(h_j-\frac{\gamma_j}{2}-\frac{\gamma_j^2}{4}\right)S_j^z+\O{\eta^3}, \label{EQ_Hup_hb}
\end{align}
where the first sum is over all $1\leq j<k\leq n$. 
Note that the first sum is also an all-to-all interaction using which we shall engineer two-qubit gates between any two spins. 
Similarly, in the ``down" subspace, i.e. the central spin is in $\ket{1}$, the effective Hamiltonian between the bath spins is
\begin{align}
	H_{\downarrow} =& \frac{1}{2} \sum_{j<k}\gamma_j\gamma_k\left(S_j^xS_k^x+S_j^yS_k^y\right)\nonumber\\
	&-\sum_{j=1}^n\left(h_j+\frac{\gamma_j}{2}-\frac{\gamma_j^2}{4}\right)S_j^z+\O{\eta^3}. \label{EQ_Hdowno_hb}
\end{align} 
In our discussion below, we shall choose $h_j = -\gamma^2_j/4$ in the system.
With this choice, the two effective Hamiltonians are different by only a sign, i.e.
\begin{align}
	H_\uparrow \approx -H_\downarrow \approx  &\sum_{j<k}\frac{\gamma_j\gamma_k}{2}\left(S_j^xS_k^x+S_j^yS_k^y\right)-\sum_{j=1}^n\frac{\gamma_j}{2} S_j^z. \label{EQ_Hup}
\end{align}
This is a key ingredient in achieving blindness later in our protocols. 
Knowing the effective Hamiltonians, we may approximate the time evolution under $H_c$ as
\begin{align}
	e^{-iH_ct} &\approx \ket{0}\bra{0}\otimes e^{-iH_{\uparrow}t} + \ket{1}\bra{1}\otimes e^{-iH_{\downarrow}t}\\
	&= e^{-i Z_0\otimes H_\uparrow t},\label{EQ_UApprox}
\end{align}
with $\ket{\alpha}\bra{\alpha}$ $(\alpha=0,1)$ being the projection operator onto the state $\ket{\alpha}$ of the central spin and $Z_0$ is the Pauli-$Z$ acting on the central spin.
Our protocol leverages the emergence of a three-spin interaction that arises naturally in the central spin settings. 
This is in contrast to many quantum gadgets simulation approaches~\cite{Kempe06,Jordan08}, where 3-local terms require substantial additional resources for implementations. 
This interaction allows the central spin to control the dynamics of the bath spins.
We now show how such a feature allows Alice and Bob to perform blind quantum simulation and, in particular, universal blind quantum computation. 

\begin{figure}[t]
	\includegraphics[width=0.45\textwidth]{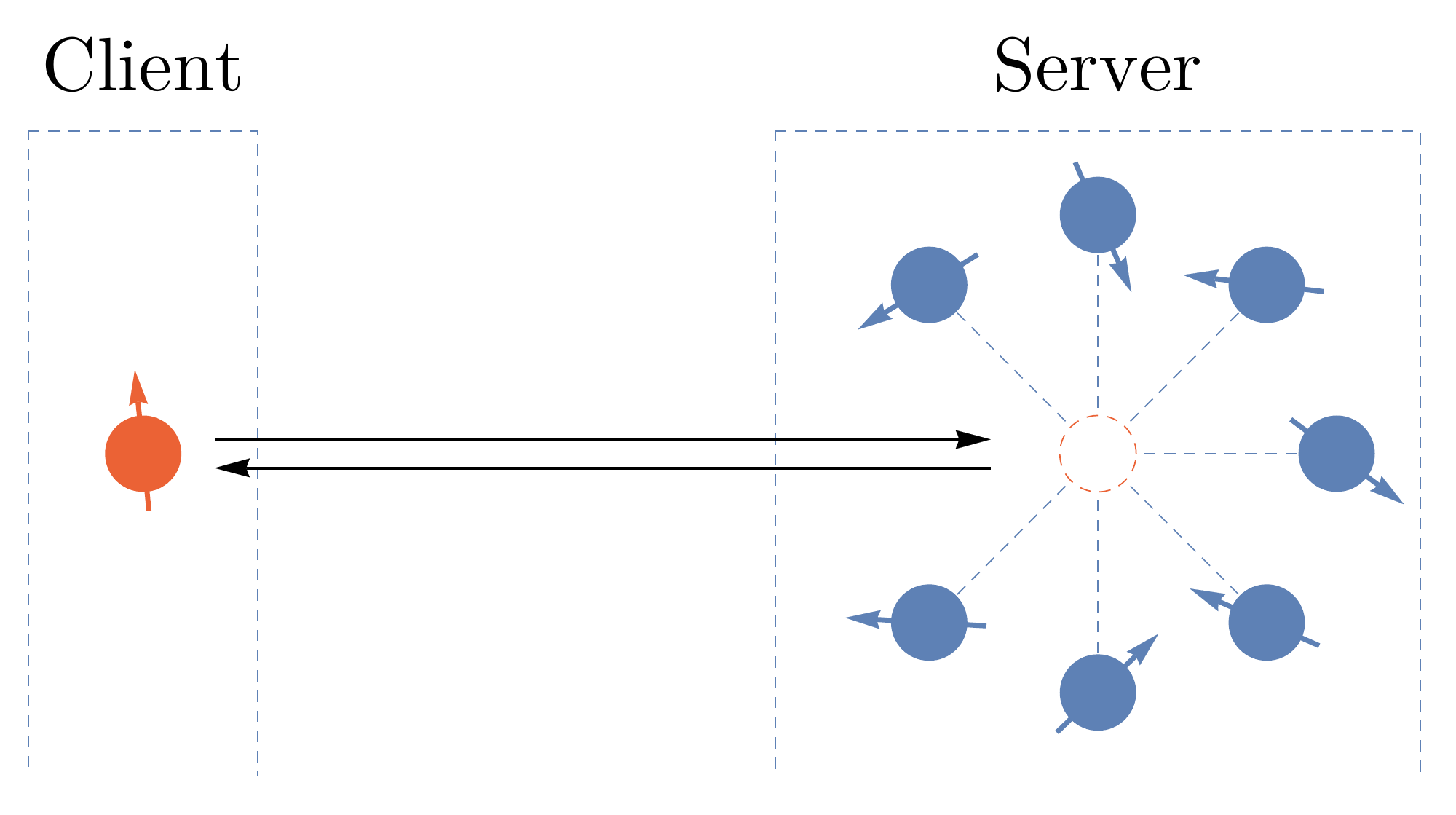}
	\caption{
		A demonstration of our blind simulation and blind computation schemes. 
		The server Bob can simulate dynamics of the central spin Hamiltonian in Eq.~\eqref{EQ_HC} where a central spin (orange) interacts with a number of bath spins (blue). 
		The bath spins can only interact with each other via the central spin.
		The effective dynamics of the bath spins is therefore encrypted in the state of the central spin, which is communicated back and forth between the server and the client.
	}
	\label{FIG_Delegate_demo}
\end{figure}


\section{Blind quantum simulation}
The three-body interaction in Eq.~\eqref{EQ_UApprox} allows a client (Alice) to perform quantum simulation on a server (Bob) such that details of the simulation are hidden from Bob.
Here we present a protocol for simulating dynamics of $n$ bath spins on the server using a single spin communicated back and forth with the client.
In our scenario, Alice has a series of $m$ Hamiltonians $H^{(k)}_{\uparrow}$ $(k=1,\dots,m)$, each of which is of the form \eqref{EQ_Hup} and is characterized by $n$ real parameters, namely the $\gamma_{j}^{(k)}$ for $j=1,\dots,n$.
Note that for each $k$, these parameters also characterize a corresponding central spin Hamiltonian $H^{(k)}_c$ as in Eq.~\eqref{EQ_HC}.
In addition to the Hamiltonians, Alice also chooses a set of $m$ constants $\{t_1,\dots,t_m\}$ and a binary string $\alpha$ of length $m$, i.e. $\alpha\in\{0,1\}^m$.
The former is to play the role of desired evolution times under the Hamiltonians of the same index and the latter is a secret key that encrypts the simulation.
The blind simulation protocol consists of $m$ iterations. In the $k$th iteration, Alice will communicate a central spin to Bob along with the classical parameters $\gamma_j^{(k)}$.
Bob then simulates evolution of the $n+1$ spins under the $H_c^{(k)}$ specified by $\gamma_j^{(k)}$ for a time $t_k$. 
Although Bob knows the evolution of the whole system, the effective time evolution of the bath spins under $H_{\uparrow}^{(k)}$ is either forward or backward in time and is encoded by the secret key $\alpha_k$ known only to Alice. 
Such a protocol for blind quantum simulation can be summarized in Protocol~\ref{PRO_SIM}.
\SetNlSkip{0.8em}
\SetInd{0.5em}{1em}
\begin{algorithm}[t]
	\caption{Blind quantum simulation} \label{PRO_SIM}
	\SetInd{1em}{1em}
	\setstretch{0}
	\DontPrintSemicolon
	\KwIn{the $n$ bath spins in a state $\ket{\psi^{(0)}}$.}
	\For{$k\in\{1,\dots,m\}$}{
		\jus{1} Alice sends Bob classical parameters $t_k$ and $\gamma_j^{(k)}$ for all $j = 1,\dots,n$.\;
		\jus{2} Alice sets the central spin to the state $\ket{\alpha_k}$ and sends to Bob.\;
		\jus{3} Bob simulates evolution of the $n+1$ spins under $H^{(k)}_c$ for time $t_k$.\; 
		\jus{4} Bob sends the central spin back to Alice.\;
	}
	\KwRet the $n$ bath spins in the final state $\ket{\psi^{(m)}}$
\end{algorithm}

Let us examine the final state of the $n$ bath spins at the end of this protocol.
Denote by $\ket{\psi^{(k)}}$ the state of the bath spins after the $k$th iteration.
By the end of line 3 in the $k$th iteration, Bob has in his possession $n+1$ spins, including the central spin sent by Alice, in the state $\ket{\alpha_{k}}\otimes\ket{\psi^{(k-1)}}$.
Bob then simulates the evolution under $H^{(k)}_c$ for the time $t_{k}$ and brings the $n+1$ spins to 
\begin{align}
	&\exp{\left(-iH^{(k)}_c t_{k}\right)}\ket{\alpha_{k}}\otimes \ket{\psi^{(k-1)}}\nonumber\\ 
	\approx& \ket{\alpha_{k}}\otimes \exp{\left(-i(-1)^{\alpha_{k}}H^{(k)}_\uparrow t_{k}\right)}\ket{\psi^{(0)}},
\end{align} 
where we have applied Eq.~\eqref{EQ_UApprox} to approximate the time evolution of the $n+1$ spins by an effective time evolution of the bath spins only. 
By induction from $k=1$ to $k=m$, the final state of the $n$ spins at the end of the protocol is
\begin{align}
	\ket{\psi^{(m)}} &= \prod_{k=1}^m \exp{\left(-i(-1)^{\alpha_{k}}H^{(k)}_\uparrow t_{k}\right)}\ket{\psi^{(0)}} \nonumber\\
	&\equiv U_m\ket{\psi^{(0)}}.\label{EQ_Unet}
\end{align}
Here the information about the effective unitary $U_m$ is partially encrypted by the key $\alpha$ known only to Alice. 

For a generic set of $H^{(k)}_c$ and $t_k$, there are $2^m$ possibilities of $U_m$ only one of which is actually implemented on the bath spins.
Therefore with a long enough key $\alpha$, Alice can be confident that Bob has almost no information about the unitary performed.
\begin{figure}[t]
	\includegraphics[width=0.5\textwidth]{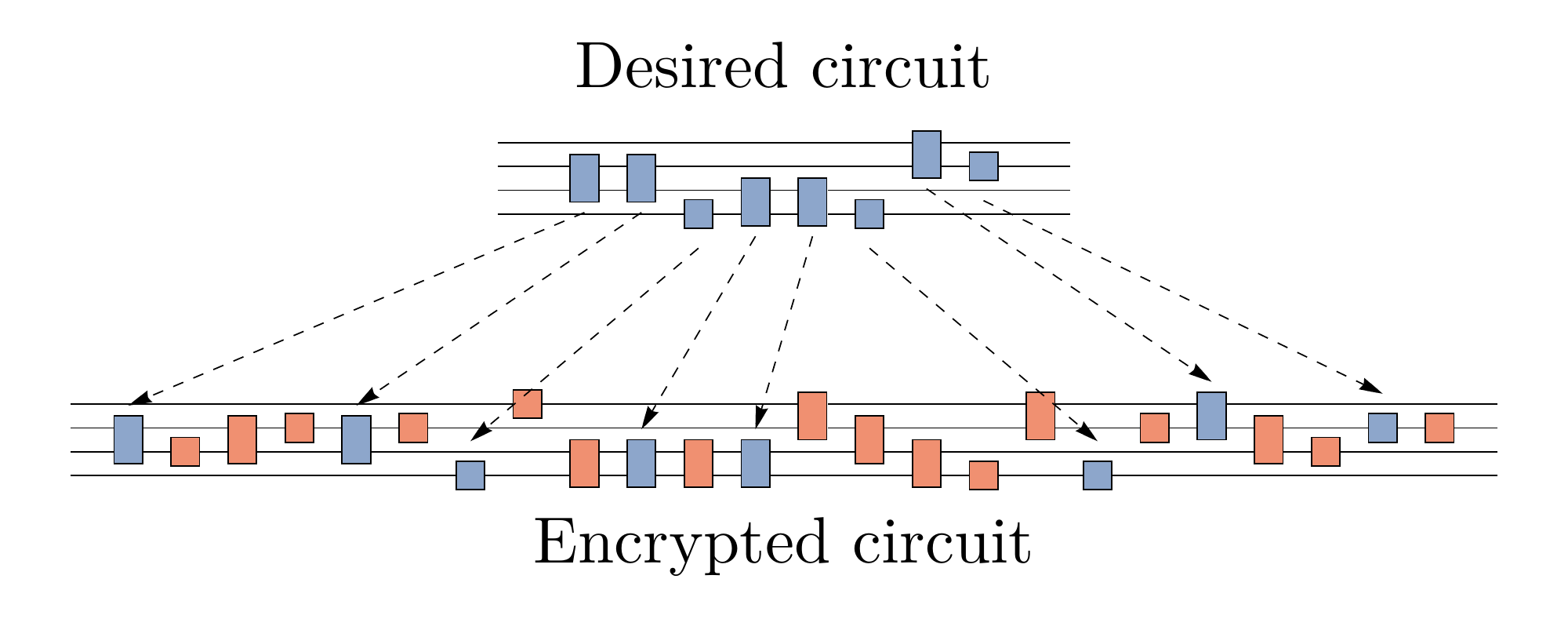}
	\caption{
		The security of Protocol~\ref{PRO_SIM_W_TRAP} and Protocol~\ref{PRO_COMP} is guaranteed by embedding the desired circuit (blue blocks) into a much longer circuit.
		The longer circuit consists of mostly honeypots (orange blocks) that do not contribute to the overall computation. 
		Instead, they serve to detect measurement attempts from the server. 
		The probability for the server to measure without being detected decays exponentially with the number of honeypots.
	}
	\label{FIG_hide_circuit}
\end{figure}
Note, however, that this simple protocol does not protect Alice's secret from a malicious Bob who tries to determine the key $\alpha$ by measuring the central spin every time Alice sends it over.

Indeed, if the protocol is followed, Bob knows the central spin can only be in one of the two orthogonal states and hence can be deterministically identified by an appropriate projective measurement.
Although Alice may not have the power to stop Bob from measuring, she can set up honeypots in the middle of the simulation to trap and abort the simulation as soon as such a malicious attempt is detected.
Using the same idea as in the BB84 quantum key distribution scheme \cite{BB84}, the set of available states of the central spin can be extended to $\left\{\ket{0},\ket{1},\ket{+},\ket{-}\right\}$ where $\ket{\pm}=\ket{0}\pm\ket{1}$ up to a normalization constant. 
Since the four states form a nonorthogonal set, it is impossible for Bob to determine with certainty which of the states is prepared by Alice. 
In particular, attempts to measure the central spin will collapse its state and therefore can be detected by Alice when the central spin is returned to her at the end of each iteration.
We present below Protocol~\ref{PRO_SIM_W_TRAP} with such honeypots.

\begin{algorithm}[t]
	\caption{Secured blind simulation} \label{PRO_SIM_W_TRAP}
	\SetInd{1em}{1em}
		\setstretch{0}
	\DontPrintSemicolon
	\KwIn{the $n$ bath spins in a state $\ket{\psi^{(0)}}$.}
	\For{$k\in\{1,\dots,m\}$}{
		\jus{1} Alice sends Bob the classical parameters $t_k$ and $\gamma_j^{(k)}$.\;
		\jus{2} Alice sets the central spin to $\ket{\alpha_k}$ and sends to Bob.\;
		\jus{3} Bob simulates evolution of the $n+1$ spins under $H^{(k)}_c$ for time $\frac{t_k}{2}$.\; 
		\jus{4} Bob returns the central spin back to Alice.\;
		\jus{5} Alice applies either a $\pi$ pulse, a $\pi/2$ pulse or identity to rotate the central spin to $\ket{\beta_k}$ and sends to Bob.\;
		\jus{6} Bob simulates evolution of the $n+1$ spins under $H^{(k)}_c$ for time $\frac{t_k}{2}$.\; 
		\jus{7} Bob sends the central spin back to Alice.\;	
		\jus{8} Alice aborts if the returned central spin is not $\ket{\beta_k}$.\;
	}
	\KwRet the bath spins in the final state $\ket{\psi^{(m)}}$.
\end{algorithm}

In addition to the key $\alpha\in\{0,1,\pm 1\}^m$, Protocol~\ref{PRO_SIM_W_TRAP} requires Alice to choose another key $\beta\in\{0,1,\pm\}^m$ of the same length $m$, with one restriction that $\beta_k = \pm$ if $\alpha_k=\mp$ for all $k$.
With this restriction, it is straightforward to verify that whenever $\alpha_k=\pm$, the net unitary applied on the bath spins in the $k$th iteration is equivalent to identity.
Such iterations therefore only play the role of flagging malicious measurement attempts and do not contribute to the overall simulation.
Note that when $\beta=\alpha\in\{0,1\}^m$, Protocol~\ref{PRO_SIM_W_TRAP} reduces to Protocol~\ref{PRO_SIM}.

%
%
\section{Universal blind computation}
So far we have shown that Alice can request a general simulation $U_m$ given by Eq.~\eqref{EQ_Unet} on $n$ bath spins without revealing her data.
But how general is $U_m$? 
In other words, what type of quantum computation Alice can achieve by simulating $U_m$?
We now show that by choosing the right parameters $\gamma_j^{(k)},t_k$ in each iteration, Alice can simulate any gate in a universal gate set and therefore is able to perform a blind simulation of an arbitrary quantum circuit.
Indeed, by turning off $\gamma_j$ for all except $j = 1$, the time evolution unitary is a local rotation of the first spin about the $z$ axis,
\begin{align}
	e^{-i \alpha Z_1 t} =e^{-i \frac{\hbar\gamma}{2}\gamma_1 t} \left(
	\begin{array}{cc}
	1 & 0 \\
	0 & e^{i\hbar\gamma_1 t} \\
	\end{array}
	\right),
\end{align}
where $Z_1$ is the Pauli-$Z$ matrix on the first qubit. 
Using this Hamiltonian, we can obtain phase-shift gates such as the $T$ gate by choosing the right evolution time $t$.
Similarly, by changing the magnetic field in Eq.~\eqref{EQ_HC} to the $x$ axis, Alice can simulate the Hadamard gate $H$.
To form a universal gate set, we still need a two-qubit gate such as the following $U_{XY}$ gate~\cite{ImamogluABDLSS99}:
\begin{align}
	U_{XY}\equiv \left(
\begin{array}{cccc}
 1 & 0 & 0 & 0 \\
 0 & \frac{1}{\sqrt{2}} & \frac{i}{\sqrt{2}} & 0 \\
 0 & \frac{i}{\sqrt{2}} & \frac{1}{\sqrt{2}} & 0 \\
 0 & 0 & 0 & 1 \\
\end{array}
\right).
\end{align}
To engineer a $U_{XY}$ gate, for example, between the first and the second qubits, we turn all $\gamma_j$ off except for $\gamma_1=\gamma_2=\gamma$. 
The effective time evolution of the bath spins is
\begin{align}
 \exp\left\{-i\left(\frac{\gamma^2}{8}\left(X_1X_2+Y_1Y_2\right)+\frac{\gamma}{2}(Z_1+Z_2)\right)t\right\},\label{EQ_UHc}
\end{align} 
where $X,Y$ are respective Pauli gates. 
With $\gamma = \frac{7}{2c}$ for some large integer $c$ and $t = \frac{7\pi}{\gamma^2}$, the above unitary reduces to the $U_{XY}$ gate.
Since the $U_{XY}$ gate and single-qubit gates form a universal gate set~\cite{ImamogluABDLSS99}, Alice can effectively perform universal quantum computation on the bath spins. 
In the following discussion, we refer to this universal gate set as $\mathcal{U}$.

Protocol~\ref{PRO_SIM_W_TRAP} can be further adapted to this situation to guarantee the security of the computation by using honeypots to detect measurement attacks.
A desired circuit $U_{m_0}$ of length $m_0$ on $n$ qubits can be embedded into a much larger circuit $U_m$ with $m\gg m_0$ that consists mostly of honeypots (Fig.~\ref{FIG_hide_circuit}).
These honeypots perform trivial operations on the qubits and serve only as detectors of malicious behaviors.
Mean while in each non-honeypot iteration of the protocol, i.e. $\alpha_k\in\{0,1\}$, Alice can choose a Hamiltonian $H^{(k)}_\uparrow$ and a time $t_k$ such that $G_k = \exp\left\{-iH^{(k)}_\uparrow t_k\right\}$ is a gate in the universal gate set $\mathcal{U}$.
Note that such a gate (or its inverse) is only implemented on the bath spins at the end of the iteration if $\beta_k \oplus \alpha_k = 0$.
On the other hand, if $\beta_k \oplus \alpha_k =1$, the two evolutions in the $k$th iteration cancel each other out and therefore only a trivial gate is implemented. 
Denote by $\omega(\alpha_k,\beta_k)$ a function of the characters $\alpha_k,\beta_k$ such that $\omega(\alpha_k,\beta_k)=1$ if $\alpha_k\oplus\beta_k=0$ and $\omega(\alpha_k,\beta_k)=0$ otherwise.
The gate sequence generated by the protocol can then be summarized by the following equation:
\begin{align}
	U_m = \prod_{k=1}^m G_{k}^{\omega(\alpha_k,\beta_k)}.\label{EQ_Ugate}
\end{align}   
Therefore the keys $\alpha,\beta$ effectively encrypt the circuit Alice implements and make the quantum computation blind.
Our protocol for universal blind quantum computation is be summarized in Protocol~\ref{PRO_COMP} below.

\begin{algorithm}[h!]
	\caption{Universal blind computation} \label{PRO_COMP}
	\SetInd{1em}{1em}
		\setstretch{0}
	\DontPrintSemicolon
	\vspace{0.1in}
	\jus{0} Alice has a circuit $U_{m_0}$ of length $m_0$ to be performed on $n$ qubits.\;
	\jus{1} Alice embeds $U_{m_0}$ into a much larger circuit $U_m$ by choosing two keys $\alpha,\beta\in\{0,1,\pm\}^m$, each of length $m\gg m_0$ such that $U_m$ in Eq.~\eqref{EQ_Ugate} reduces to $U_{m_0}$.\;
	\jus{2} Alice and Bob perform Protocol~\ref{PRO_SIM_W_TRAP} to implement $U_m$ on $\ket{\psi}$.\;
\end{algorithm}

\section{Quantum verification}
Blindness 
allows the client Alice to not only hide the computation from the server Bob but also to verify if Bob performs the correct computation.
Indeed, Alice can verify Bob by simply requesting quantum circuits that have outcomes that can be classically \emph{verified}. 
By the definition of blind computation used in this work, Bob has no information about what circuits are being implemented, and the only way he can return the correct output to Alice is to perform the exact simulation sequence as instructed.
For example, Alice can ask Bob to initialize the $n$ bath spins in a product state such that some of the spins are ``up" and some are ``down", and use Protocol~\ref{PRO_COMP} to simulate a  sequence of SWAP gates between the bath spins known only to Alice.
The final state is a permutation of the initial state and is known only to Alice. 
Bob has to find the final state and the only way he can pass with certainty is to correctly perform the computation.

Since both the initial state and the final state are fully separable, the permutation circuit is essentially classical. 
Stabilizer circuits~\cite{Bennett96,Gottesman96}, on the other hand, can perform nontrivial quantum operations, such as quantum teleportation~\cite{Bennett93} and preparation of highly entangled states.
They consist of only Clifford gates and can be simulated efficiently on a classical computer~\cite{Aaronson04}. 
Therefore by requesting simulation of an arbitrary stabilizer circuits, Alice can also efficiently verify quantumness of the server.
 
Alice can also take a step further to verify even the quantum computing power of the server by requesting a quantum circuit that is known to solve a problem faster than classical algorithms.
For example, in the Simon's problem \cite{Simon97}, a function $f:\{0,1\}^n\rightarrow\{0,1\}^n$ is promised to satisfy that $f(x)=f(y)$ if and only if $x=y$ or $x\oplus y = s$ for all $x,y\in\{0,1\}^n$ and a fixed string $s\in\{0,1\}^n$.
To find $s$, classical algorithms require at least $\Omega(2^{n/2})$ queries to the function $f$ while the quantum Simon's algorithm can solve the problem using only $\O{n}$ queries.
Using Protocol~\ref{PRO_COMP}, Alice can simulate a quantum circuit corresponding to a secret string $s$. 
She then asks Bob to measure the output and announce the measured string. 
If Bob is able to answer correctly what $s$ is for large enough $n$, Alice can be confident that the server Bob has access to at least a BQP machine.


\section{Outlook}
Here we have shown how to implement an arbitrary circuit controlled by a single spin.
This enables us to define several blind computing protocols that can be a powerful test of computing power in quantum simulators. 
However, we have not yet developed natural observables whose, e.g., distribution function is distinctly different given classical versus quantum computational power.
We consider this an intriguing direction for future research. 

\begin{acknowledgments}
We thank S. -H Hung, B. Lackey, R. Matthew, and Y. Wang for helpful discussions. This research was supported in part by the NSF funded Physics Frontier Center at the Joint Quantum Institute and the Army Research Laboratory's CDQI.
\end{acknowledgments}

\bibliographystyle{apsrev4-1}
\bibliography{blind-comp}

\appendix
\section{Schrieffer-Wolff approximation}
\label{APP_SW}
In this section, we show how the Schrieffer-Wolff transformation \cite{Bravyi11} reduces the $(n+1)$-spin Hamiltonian $H_c$ in Eq.~\eqref{EQ_HC} to the effective Hamiltonians in Eq.\eqref{EQ_Hup_hb} and Eq.~\eqref{EQ_Hdowno_hb}.
We first divide the Hamiltonian $H_c$ into three parts, namely
\begin{align}
H_0 &= -h_0 S_0^z,\\
H_d &= -h_b \sum_{j=1}^n S_j^z + \sum_{j=1}^n \gamma_j S_0^zS_j^z,\\
H_{od} &=  \sum_{j=1}^n \gamma_j \left(S_0^xS_j^x+S_0^yS_j^y\right)\\
&=  \frac12\sum_{j=1}^n \gamma_j \left(S_0^+S_j^-+S_0^-S_j^+\right),
\end{align}
where $S^\pm = S^x\pm i S^y$.
Since $h_0\gg n\gamma\equiv \eta$ with $\gamma$ being the average amplitude of the $\gamma_j$, we treat $H_0$ as the unperturbed Hamiltonian and $H_{d}+H_{od}$ as the perturbation in our approximation.
The eigenspace of $H_0$ is well separated into two subspaces corresponding to the two eigenvalues $\pm h_0 \frac{\hbar}{2}$.
Note that the Hamiltonian $H_d$ is diagonal in this block representation of the Hilbert space while the Hamiltonian $H_{od}$ is off-diagonal and hence induces interaction between the two subspaces.
It is this off-diagonal part of the Hamiltonian that gives rise to the effective interaction between the bath spins, despite them being not directly coupled in the original Hamiltonian $H_c$.

The idea of Schrieffer-Wolff approximation is to block diagonalize the Hamiltonian in the basis of $H_0$.
Such a block diagonalization can be achieved using a unitary $U = e^{T}$ with $T$ being an anti-Hermitian operator.
In Ref.~\cite{Bravyi11}, the operator $T$ is expanded using a Taylor series, i.e. $T = \sum_{k=1}^\infty T_k \eta^k$ in terms of the perturbation order $\eta = n\gamma$.
In the following discussion, we shall absorb the order $\eta^k$ into $T_k$.
The effective Hamiltonian in the low energy subspace of $H_0$ is given by \cite{Bravyi11}
\begin{align}
H_\uparrow =& H_0 P_0 +P_0(H_d+H_{od})P_0 \nonumber\\
&+ \frac12 P_0 \comm{T_1,H_{od}}P_0 +\O{\eta^3}, \label{EQ_Hup_form}
\end{align}
where $P_0=\ket{0}\bra{0}$ is the projector onto the lower energy subspace of $H_0$.
It is straightforward to calculate the first two terms,
\begin{align}
H_0P_0 &= -\frac12 h_0,\label{APP_EQ_1st_term}\\
P_0(H_d+H_{od})P_0 &= P_0H_{d}P_0 \\
&=  -h_b \sum_{j=1}^n S_j^z +\frac{1}{2}  \sum_{j=1}^n \gamma_j S_j^z.\label{APP_EQ_2nd_term}
\end{align}
To calculate the third term, we use the formula given in Section 3.2 in Ref.~\cite{Bravyi11} to first find the operator $T_1$,
\begin{align}
 T_1  =&  \frac12\frac{\bra{0}S_0^+\ket{1}}{E_+-E_-}\ket{0}\bra{1}\sum_{j=1}^n \gamma_j S_j^-\nonumber\\
&+ \frac12\frac{\bra{1}S_0^-\ket{0}}{E_--E_+}\ket{0}\bra{1}\sum_{j=1}^n \gamma_j S_j^+\\
=&  -\frac1{2h}\sum_{j=1}^n \gamma_j \left(S_0^+S_j^--S_0^-S_j^+\right).
\end{align}
where $E_\pm = \mp h_0/2$ are the eigenvalues of $H_0$.
Thus the third term in Eq.~\eqref{EQ_Hup_form} is
\begin{widetext}
\begin{align}
\frac12 P_0 [T_1,H_{od}]P_0
&=-\frac{1}{8h_0} P_0\sum_{j,k} \gamma_j\gamma_k\left[S_0^+S_j^--S_0^-S_j^+,S_0^+S_k^-+S_0^-S_k^+\right]P_0\\
&=-\frac{}{2h_0}\sum_{j< k} \gamma_j\gamma_k\left( S_j^xS_k^x+S_j^yS_k^y\right)+\frac{1}{4 h} \sum_{j} \gamma_j^2 S_j^z-\frac{1}{8h}\sum_{j}\gamma_j^2.\label{APP_EQ_3rd_term}
\end{align}
\end{widetext}
Here we have omitted the straightforward simplification from the first to the second line.
Combining Eq.~\eqref{APP_EQ_1st_term}, Eq.~\eqref{APP_EQ_2nd_term} and Eq.~\eqref{APP_EQ_3rd_term} we have the effective Hamiltonian between the bath spins in the low energy subspace (up to a constant),
\begin{align}
H_{\uparrow} =& -\frac{1}{2h_0} \sum_{j<k}\gamma_j\gamma_k\left(S_j^xS_k^x+S_j^yS_k^y\right)\nonumber\\
&-\sum_{j=1}^n\left(h_j-\frac{\gamma_j}{2}-\frac{\gamma_j^2}{4h_0}\right)S_j^z+\O{\eta^3}. \label{APP_EQ_Hup_hb}
\end{align}
We note that this is the effective Hamiltonian in a frame rotated by $e^T$.
However, in our scenario, $e^T = \mathbb{I}+\O{\eta^2}$ is approximately identity.
Therefore after rotating back to the laboratory frame and taking only the leading orders, the effective Hamiltonian between the bath spins in the low energy subspace is still given by Eq.~\eqref{APP_EQ_Hup_hb}.

The effective Hamiltonian in the high energy subspace, i.e. the space corresponding to the central spin being $\ket{1}$, can be found using the exact same steps. 
However, instead of using the projector $P_0$, we project onto the high energy subspace $Q_0 = \mathbb{I}-P_0$.
In the end, we have a system in which the effective Hamiltonian can be switched between $H_\uparrow$ and $H_\downarrow$ by controlling the state of the central spin.
Such a feature is captured by an unitary on the whole system
\begin{align}
	e^{-iH_ct} &\approx \ket{0}\bra{0}\otimes e^{-iH_{\uparrow}t} + \ket{1}\bra{1}\otimes e^{-iH_{\downarrow}t}.
\end{align}

\end{document}